\documentclass[twocolumn,showpacs,preprintnumbers,amsmath,amssymb]{revtex4}

\usepackage{graphicx}
\usepackage{dcolumn}
\usepackage{bm}

\begin{document}

\preprint{}

\title{Optimal state in the Knill-Laflamme-Milburn scheme of linear optical teleportation}

\author{Andrzej Grudka}

\affiliation{Faculty of Physics, Adam Mickiewicz University,
Umultowska 85, 61-614 Pozna\'{n}, Poland}

\author{Joanna Mod{\l}awska}

\affiliation{Faculty of Physics, Adam Mickiewicz University,
Umultowska 85, 61-614 Pozna\'{n}, Poland}

\date{\today}

\begin{abstract}
We discuss some properties of the Knill-Laflamme-Milburn scheme for quantum teleportation with both maximally and nonmaximally entangled states. We derive the error correction scheme when one performs teleportation with nonmaximally entangled states and we find the probability for perfect teleportation. We show that the maximally entangled state is optimal in such a case. We also show how the error correction scheme can be implemented experimentally when one uses polarization encoding.
\end{abstract}

\maketitle

\section{Introduction}

A great progress in linear-optical quantum computation \cite{Kok} was due to Knill, Laflamme, and Milburn (KLM) who derived a scheme for efficient computation \cite{Knill}. They showed how to perform computation with single photon sources, linear optical elements (i.e., phase-shifters, beam-splitters), and projective measurements in the Fock basis (i.e., photodetectors able to discriminate the number of photons). One of the crucial ingredients of their scheme is a protocol for linear optical teleportation. They showed that one can obtain a probability of successful teleportation arbitrarily close to 1. More presicesly they use certain $n$-photon maximally entangled state to teleport a qubit which is a superposition of the vacuum and one-photon states. The probability of successful teleportation is equal to $1-\frac{1}{n+1}$. What is important in the KLM approach is the fact that if the teleportation is successful then the fidelity of the teleported qubit is equal to 1. Hence, the average fidelity scales as $1-O(\frac{1}{n})$. Their protocol can be combined with the Gottesman-Chuang protocol \cite{Gottesman} to obtain a two-photon controlled-$Z$ gate. Spedalieri \emph{et al.} \cite{Spedalieri} generalized the KLM protocol to polarization encoding. Franson \emph{et al.} \cite{Franson} presented a different approach to quantum teleportation. They relaxed the condition of perfect fidelity and concentrated only on average fidelity. Thus, they assumed that teleportation is always successful and tried to maximize the average fidelity.  They showed that one can obtain an average fidelity for the teleported qubit which scales as $1-O\left(\frac{1}{n^2}\right)$ when one uses a carefully chosen $n$-photon nonmaximally entangled state. The latter result is interesting because usually maximally entangled states are assumed to be better for information-theoretic tasks. However, it is not difficult to see why a nonmaximally entangled state performs better here. In the KLM scheme one cannot perform a measurement in the generalized Bell basis \cite{Lutkenhaus}, i.e., one cannot project onto a state which has an indefinite number of photons. Instead, with probability $\frac{1}{n+1}$ one registers $0$ or $n+1$ photons and, thus, destroys the state of the qubit to be teleported. Franson \emph{et al.} lowered this probability by a careful choice of the entangled state. However, they had to pay a price. Their scheme introduced some small error in the teleported state. In Ref. \cite{Kok} it was stated that \emph{this makes error correction much harder}.

In this paper we present an optimal scheme of error correction and find the probability of obtaining a fidelity equal to $1$ for the teleported state. This provides a link between both schemes. Moreover, we show that if one wants to obtain perfect fidelity of the teleported state then the maximally entangled state is optimal. The paper is organized as follows. In Sec. II we describe our error correction scheme. In Sec. III we prove optimality of the KLM state. In Sec. IV we show how the error correction can be performed experimentally for polarization encoding. We conclude in Sec. V.

\section{Error correction for teleportation with nonmaximally entangled states}

The entangled state in the generalization of the KLM protocol for linear optical teleportation has the form
\begin{equation}
|t_{n}\rangle=\sum_{i=0}^{n}c_{i}|1\rangle^{i}|0\rangle^{n-i}|0\rangle^{i}|1\rangle^{n-i},
\label{eq:1}
\end{equation}
where $|k\rangle^{i}$ stands for $|k\rangle_{1}|k\rangle_{2}...|k\rangle_{i}$, i.e., $k$ photons in each of the subsequent modes. To teleport a qubit in the state $|\psi\rangle=\alpha|0\rangle+\beta|1\rangle$ one applies the $n+1$-point quantum Fourier transform to the input mode and to the first $n$ modes of the state $|t_{n}\rangle$, which is given by
\begin{equation}
F_{n} (a_{k}^{\dagger})=\frac{1}{\sqrt{n+1}} \sum_{l_{k}=0}^{n} \omega^{k l_{k}} a_{l_{k}}^{\dagger}.
\end{equation}
In the equation above, $a_{k}^{\dagger}$ is the creation operator for a photon in the $k$th mode  and $\omega = e^{i 2 \pi/(n+1)}$. In the next step one measures the total number of photons in these $n+1$ modes. If $m$ photons are detected in total then the modified state of the qubit is found in the mode $n+m$. More precisely, after phase correction one has the following state in the $n+m$th mode
\begin{equation}
|\psi_{m}\rangle=\frac{1}{\sqrt{p(m)}}(\alpha c_{m}|0\rangle+\beta c_{m-1}|1\rangle),
\label{eq:2}
\end{equation}
where $p(m)$ is the probability of detecting $m$ photons and is given by
\begin{equation}
p(m)=|\alpha c_{m}|^2+|\beta c_{m-1}|^2.
\label{eq:3}
\end{equation}
Let us show how one can perform error correction to retrieve the original state of the qubit. Since the coefficients $c_{i}$ are known, the best that one can do is to perform a generalized measurement on the photon in the $n+m$th mode \cite{Gisin}. This measurement is given by the following pair of Kraus operators:
\begin{eqnarray}
E_{S}=\frac{c_{m-1}}{c_{m}}|0\rangle \langle0|+|1\rangle \langle1|, \nonumber\\
E_{F}=\sqrt{1-\left|\frac{c_{m-1}}{c_{m}}\right|^2}|0\rangle \langle0|
\label{eq:4}
\end{eqnarray}
for $|c_{m-1}|^2 \leq |c_{m}|^2$
and
\begin{eqnarray}
E_{S}=|0\rangle \langle0|+\frac{c_{m}}{c_{m-1}}|1\rangle \langle1|, \nonumber\\
E_{F}=\sqrt{1-\left|\frac{c_{m}}{c_{m-1}}\right|^2}|1\rangle \langle1|
\label{eq:5}
\end{eqnarray}
for $|c_{m-1}|^2 > |c_{m}|^2$. If $S$ is obtained as the result of the measurement then the post-measurement state of the qubit is $\alpha|0\rangle+\beta|1\rangle$ and one can see that error correction succeeded. It is of great importance that the error correction was made after teleportation rather than before. The probability of this event is:
\begin{equation}
p(S|m)=\langle\psi_{m}|E_{S}^{\dagger}E_{S}|\psi_{m}\rangle=\frac{|c_{m}^{<}|^2}{p(m)},
\label{eq:6}
\end{equation}
where $c_{m}^{<}=\text{min}\{|c_{m-1}|, |c_{m}|\}$. Hence, the joint probability of detecting $m$ photons and successful error correction is
\begin{equation}
p(S,m)=p(S|m)p(m)=|c_{m}^{<}|^2.
\label{eq:7}
\end{equation}
One should stress that this probability is independent of the state of the teleported qubit. Otherwise one would gain information on the state, which is impossible if the qubit is teleported faithfully.

We are now ready to find the total probability of successful teleportation, i.e., $p(S)=\sum_{m=0}^{n}p(S,m)$. It should be stressed that by successful teleportation we mean teleportation with unit fidelity. In Fig. 1 we present a possible choice of the coefficients $|c_{m}|^{2}$ of the entangled state. As can be seen, this sequence can be increasing as well as decreasing. Let us first consider what happens if the sequence increases, i.e., $|c_{m-2}|^2<|c_{m-1}|^2<|c_{m}|^2$. With the help of Eq.~(\ref{eq:7}) we find that the probability of detecting $m-1$ $(m)$ photons and successful error correction is $|c_{m-2}|^{2}$ $(|c_{m-1}|^{2})$ and, hence, $p(S,m-1)+p(S,m)=|c_{m-2}|^2+|c_{m-1}|^2$. On the other hand, if the sequence decreases, i.e., $|c_
{m-2}|^{2}>|c_{m-1}|^{2}>|c_{m}|^{2}$, we find that $p(S,m-1)+p(S,m)=|c_{m-1}|^2+|c_{m}|^2$. In general, we have
\begin{equation}
\sum_{m=m_{1}+1}^{m_{2}} p(S,m)=\sum_{m=m_{1}}^{m_{2}-1} |c_{m}|^2
\label{eq:8}
\end{equation}
for an increasing sequence and
\begin{equation}
\sum_{m=m_{1}+1}^{m_{2}} p(S,m)=\sum_{m=m_{1}+1}^{m_{2}} |c_{m}|^2
\label{eq:9}
\end{equation}
for a decreasing one.

\begin{figure}
\includegraphics [width=7truecm]{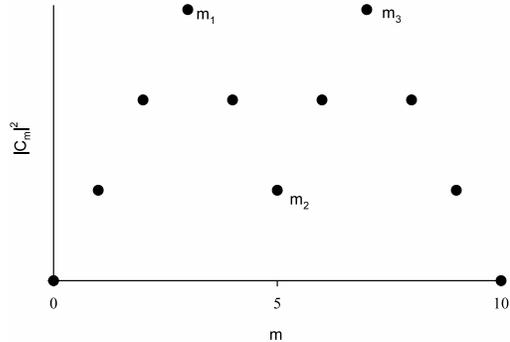}
\caption{\label{fig:1} Exemplary sequence $|c_{m}|^2$.}
\end{figure}

We now investigate in more detail what happens in the vicinity of a maximum and of a minimum of the sequence. Let us begin with a maximum, i.e., we consider the relation $|c_{m-2}|^2=|c_{m}|^2<|c_{m-1}|^2$ between the coefficients appearing in the sequence. We find that 
\begin{equation}
p(S,m-1)+p(S,m)=|c_{m-2}|^{2}+|c_{m}|^{2}.
\label{eq:10}
\end{equation}
We see that a maximum never appears in the expression for the total probability of successful teleportation. On the other hand, if the sequence has a minimum rather than a maximum, i.e., $|c_{m-2}|^2=|c_{m}|^2>|c_{m-1}|^2$, we obtain
\begin{equation}
p(S,m-1)+p(S,m)=2|c_{m-1}|^{2}.
\label{eq:11}
\end{equation}
Thus, if the sequence has a minimum for $m \neq 0$ or $m \neq n$, its value appears doubled in the expression for the probability of successful teleportation. It should be stressed at this point that there is a difference in treating a maximum or a minimum for $m = 0$ or $m = n$. In such cases, the maximum is not counted when calculating the probability of successful teleportation and the minimum is counted only once. 

It is not difficult now to derive the analytic formula for the total probability of successful teleportation. Let us suppose that our sequence has minima for $m=0$, $m=m_{2}$, $m=m_{4}$,..., and maxima for $m=m_{1}$, $m=m_{3}$,...,  and so on. We can divide it into parts which increase, i.e., $0 < m \leq m_{1}$, $m_{2}< m \leq m_{3}$,..., and parts which decrease $m_{1} < m \leq m_{2}$, $m_{3} < m \leq m_{4}$,..., and so on. We use several times Eq.~(\ref{eq:8}) in the first case (increasing sequence), and Eq.~(\ref{eq:9}) in the second case (decreasing sequence). Using the fact that $\sum_{m=0}^{n}|c_{m}|^2=1$, the formula for the total probability of successful teleportation takes the form
\begin{equation}
p(S)=1 - \sum_{\text{max}} |c_{m}|^2 +\sum_{\text{min}, m \neq 0, m \neq n} |c_{m}|^2,
\label{eq:12}
\end{equation}
where the first sum is over all $m$ for which the sequence has maxima and the second sum is over all $m$ for which the sequence has minima except $m=0$ and $m=n$ if such exist.

\section{Optimality of the KLM state}

With the derived formula we can now ask the following question: Which state [among the ones described by Eq.~(\ref{eq:1})] is optimal for successful teleportation, i.e., which one gives the maximal probability of obtaining the original state of the teleported qubit? To answer
this question we must investigate the relation between maxima and minima in the sequence $\{|c_{i}|^2\}$. Let the largest maximum be $|c_{M}|^2$. Since 
\begin{equation}
1=\sum_{m=0}^{n} |c_{m}|^2 < \sum_{m=0}^{n} |c_{M}|^2=(n+1)|c_{M}|^2,
\label{eq:13}
\end{equation}
we have $|c_{M}|^2 > \frac{1}{n+1}$. Let us now rewrite our formula in the following way
\begin{equation}
p(S)=1 - |c_{M}|^2 -  \left( \sum_{\text{max}, m \neq M} |c_{m}|^2 -\sum_{\text{min}, m \neq 0, m \neq n} |c_{m}|^2 \right),
\label{eq:14}
\end{equation}
where now we do not count in the first sum the maximum at $m=M$.
Since, as already mentioned, we do not count the minima for  $m=0$ and $m=n$, we have the same number of maxima and minima for $m<M$. Now each  maximum is greater than the neighboring minimum. The same reasoning holds also for $m>M$. Hence, the following inequality holds: 
\begin{equation}
\sum_{\text{max}, m \neq M} |c_{m}|^2 -\sum_{\text{min}, m \neq 0, m \neq n} |c_{m}|^2 \geq 0.
\label{eq:15}
\end{equation}
We can now see that for each sequence which has at least one maximum, the probability of successful teleportation satisfies the inequality $p(S)<1-\frac{1}{n+1}$. We conclude that the optimal state is the one which has the squared moduli of all coefficients equal, i.e., $|c_{m}|^2 = \frac{1}{n+1}$.

\section{Implementation of error correction with polarization encoding}

Let us now describe how the proposed method of error correction can be implemented experimentally with polarization encoding \cite{Spedalieri}. Now, instead of the state of Eq.~(\ref{eq:1}), we have the state
\begin{equation}
|t_{n}\rangle=\sum_{i=0}^{n}c_{i}|V\rangle^{i}|H\rangle^{n-i}|H\rangle^{i}|V\rangle^{n-i},
\label{eq:16}
\end{equation}
and the state to be teleported is $|\psi\rangle=\alpha|H\rangle+\beta|V\rangle$, where $|H\rangle$ and $|V\rangle$ stand for horizontal and vertical polarization, respectively. In order to perform teleportation we apply the $n+1$-point quantum Fourier transform to the input mode and the first $n$ modes of the state $|t_{n}\rangle$, i.e.,
\begin{eqnarray}
F_{n} (h_{k}^{\dagger})=\frac{1}{\sqrt{n+1}} \sum_{l_{k}=0}^{n} \omega^{k l_{k}} h_{l_{k}}^{\dagger},\nonumber\\
F_{n} (v_{k}^{\dagger})=\frac{1}{\sqrt{n+1}} \sum_{l_{k}=0}^{n} \omega^{k l_{k}} v_{l_{k}}^{\dagger},
\end{eqnarray}
where $h_{k}^{\dagger}$ and $v_{k}^{\dagger}$ are creation operators for horizontally and vertically polarized photons in the $k$th mode, respectively. In the next step, we measure the number of photons in each polarization in these modes. If  we find $m$ horizontally polarized photons and $n-m+1$ vertically polarized photons then the modified state of the qubit is teleported to the $m+n$th mode and it is now (after phase correction)
\begin{equation}
|\psi_{m}\rangle=\frac{1}{\sqrt{p(m)}}(\alpha c_{m}|H\rangle+\beta c_{m-1}|V\rangle),
\label{eq:18}
\end{equation}
where $p(m)$ is given by Eq.~(\ref{eq:3}). We can perform error correction as was done it in Sec. II. We perform a generalized measurement given by Kraus operators of Eqs.~(\ref{eq:4}) and (\ref{eq:5}) with $|0\rangle \langle 0|$ and $|1\rangle \langle 1|$ replaced by $|H\rangle \langle H|$ and $|V\rangle \langle V|$, respectively. Since we have now polarization encoding rather than photon-number encoding, this error correction can be easily implemented experimentally. 

\begin{figure}
\includegraphics [width=6truecm]{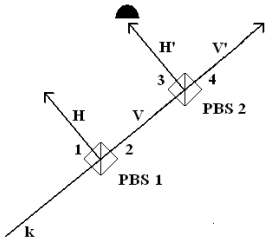}
\caption{\label{fig:2} Experimental setup for error correction with polarization encoding.}
\end{figure}

In Fig. 2 we present an experimental setup which performs this task. First, the photon from the $k$th mode enters one of two input ports of a polarizing beam splitter while the other port is left empty. The polarizing beam splitter reflects horizontally polarized photons and transmits vertically polarized photons.  Let us suppose that $|c_{m}|^2<|c_{m-1}|^2$. Another polarizing beam splitter is added in mode $2$. It is rotated by an angle $\theta$, where
\begin{equation}
\cos{\theta}=\left|\frac{c_{m}}{c_{m-1}}\right|.
\end{equation}
Thus, it reflects photons of polarization
\begin{equation}
|H' \rangle = - \sin{\theta}|V\rangle + \cos{\theta}|H\rangle
\end{equation}
and transmits photons of polarization
\begin{equation}
|V' \rangle = \cos{\theta}|V\rangle + \sin{\theta}|H\rangle.
\end{equation}
The state of the photon after the second polarizing beam splitter is
\begin{eqnarray}
|\psi_{m}\rangle=\frac{1}{\sqrt{p(m)}}(\alpha c_{m}|H\rangle_{1}+\nonumber\\
+\beta c_{m-1}\cos{\theta}|V'\rangle_{4}-\beta c_{m-1} \sin{\theta}|H'\rangle_{3}).
\end{eqnarray}
Finally, there is a detector in one of two output ports of the polarizing beam splitter. If it registers a photon then the state of a qubit is detroyed. However, if it does not register a photon then the qubit is found in the state
\begin{equation}
\alpha |H\rangle_{1} + \beta |V' \rangle_{4}.
\end{equation}
If one performs a rotation of polarization state of the photon in mode $4$, i.e., $|V' \rangle_{4} \rightarrow |V \rangle_{4}$, one recovers the original qubit state, which has therefore been faithfully teleported. If $|c_{m}|^2>|c_{m-1}|^2$, a polarizing beam splitter is added in mode $1$ rather than in mode $2$.

\section{Conclusions}

We have shown how one can perform error correction for the KLM scheme of teleportation with nonmaximally multimode entanglement states. We have derived a formula for probability of successful teleportation, i.e., teleportation with unit fidelity. We have shown that the maximally entangled state is optimal in such a case. However, this does not contradict that there may be certain information-theoretic tasks for which nonmaximally entangled states are better, e.g. the ones considered by Franson \emph{et al.} \cite{Franson}.  Indeed, after the completion of this work we were able to show that nonmaximally entangled states are optimal for multiple linear optical teleportation \cite{Modlawska}. Finally we have shown how to perform error correction experimentally.

\begin{acknowledgments} The authors acknowledge discussions with Rafa{\l} Demkowicz-Dobrza\'{n}ski, Adam Miranowicz, and Marco Piani. One of the authors (A.G.) was supported by the State Committee for Scientific Research Grant No. P03B 014 30.
\end{acknowledgments}


\begin{thebibliography}{8}
\expandafter\ifx\csname natexlab\endcsname\relax\def\natexlab#1{#1}\fi
\expandafter\ifx\csname bibnamefont\endcsname\relax
  \def\bibnamefont#1{#1}\fi
\expandafter\ifx\csname bibfnamefont\endcsname\relax
  \def\bibfnamefont#1{#1}\fi
\expandafter\ifx\csname citenamefont\endcsname\relax
  \def\citenamefont#1{#1}\fi
\expandafter\ifx\csname url\endcsname\relax
  \def\url#1{\texttt{#1}}\fi
\expandafter\ifx\csname urlprefix\endcsname\relax\def\urlprefix{URL }\fi
\providecommand{\bibinfo}[2]{#2}
\providecommand{\eprint}[2][]{\url{#2}}

\bibitem[{\citenamefont{Kok et~al.}(2007)\citenamefont{Kok, Munro, Nemoto,
  Ralph, Dowling, and Milburn}}]{Kok}
\bibinfo{author}{\bibfnamefont{P.}~\bibnamefont{Kok}},
  \bibinfo{author}{\bibfnamefont{W.~J.} \bibnamefont{Munro}},
  \bibinfo{author}{\bibfnamefont{K.}~\bibnamefont{Nemoto}},
  \bibinfo{author}{\bibfnamefont{T.~C.} \bibnamefont{Ralph}},
  \bibinfo{author}{\bibfnamefont{J.~P.} \bibnamefont{Dowling}},
  \bibnamefont{and} \bibinfo{author}{\bibfnamefont{G.~J.}
  \bibnamefont{Milburn}}, \bibinfo{journal}{Rev. Mod. Phys.}
  \textbf{\bibinfo{volume}{79}}, \bibinfo{pages}{135} (\bibinfo{year}{2007}).

\bibitem[{\citenamefont{Knill et~al.}(2001)\citenamefont{Knill, Laflamme, and
  Milburn}}]{Knill}
\bibinfo{author}{\bibfnamefont{E.}~\bibnamefont{Knill}},
  \bibinfo{author}{\bibfnamefont{R.}~\bibnamefont{Laflamme}}, \bibnamefont{and}
  \bibinfo{author}{\bibfnamefont{G.~J.} \bibnamefont{Milburn}},
  \bibinfo{journal}{Nature} \textbf{\bibinfo{volume}{409}}, \bibinfo{pages}{46}
  (\bibinfo{year}{2001}).

\bibitem[{\citenamefont{Gottesman and Chuang}(1999)}]{Gottesman}
\bibinfo{author}{\bibfnamefont{D.}~\bibnamefont{Gottesman}} \bibnamefont{and}
  \bibinfo{author}{\bibfnamefont{I.~L.} \bibnamefont{Chuang}},
  \bibinfo{journal}{Nature} \textbf{\bibinfo{volume}{402}},
  \bibinfo{pages}{390} (\bibinfo{year}{1999}).

\bibitem[{\citenamefont{Spedalieri et~al.}(2006)\citenamefont{Spedalieri, Lee,
  and Dowling}}]{Spedalieri}
\bibinfo{author}{\bibfnamefont{F.~M.} \bibnamefont{Spedalieri}},
  \bibinfo{author}{\bibfnamefont{H.}~\bibnamefont{Lee}}, \bibnamefont{and}
  \bibinfo{author}{\bibfnamefont{J.~P.} \bibnamefont{Dowling}},
  \bibinfo{journal}{Phys. Rev. A} \textbf{\bibinfo{volume}{73}},
  \bibinfo{pages}{012334} (\bibinfo{year}{2006}).

\bibitem[{\citenamefont{Franson et~al.}(2002)\citenamefont{Franson, Donegan,
  Fitch, Jacobs, and Pittman}}]{Franson}
\bibinfo{author}{\bibfnamefont{J.~D.} \bibnamefont{Franson}},
  \bibinfo{author}{\bibfnamefont{M.~M.} \bibnamefont{Donegan}},
  \bibinfo{author}{\bibfnamefont{M.~J.} \bibnamefont{Fitch}},
  \bibinfo{author}{\bibfnamefont{B.~C.} \bibnamefont{Jacobs}},
  \bibnamefont{and} \bibinfo{author}{\bibfnamefont{T.~B.}
  \bibnamefont{Pittman}}, \bibinfo{journal}{Phys. Rev. Lett.}
  \textbf{\bibinfo{volume}{89}}, \bibinfo{pages}{137901}
  (\bibinfo{year}{2002}).

\bibitem[{\citenamefont{Lutkenhaus et~al.}(1999)\citenamefont{Lutkenhaus,
  Calsamiglia, and Suominen}}]{Lutkenhaus}
\bibinfo{author}{\bibfnamefont{N.}~\bibnamefont{Lutkenhaus}},
  \bibinfo{author}{\bibfnamefont{J.}~\bibnamefont{Calsamiglia}},
  \bibnamefont{and} \bibinfo{author}{\bibfnamefont{K.~A.}
  \bibnamefont{Suominen}}, \bibinfo{journal}{Phys. Rev. A}
  \textbf{\bibinfo{volume}{59}}, \bibinfo{pages}{3295} (\bibinfo{year}{1999}).

\bibitem[{\citenamefont{Gisin}(1996)}]{Gisin}
\bibinfo{author}{\bibfnamefont{N.}~\bibnamefont{Gisin}},
  \bibinfo{journal}{Phys. Lett. A} \textbf{\bibinfo{volume}{210}},
  \bibinfo{pages}{151} (\bibinfo{year}{1996}).

\bibitem[{\citenamefont{Mod{\l}awska and Grudka}(2007)}]{Modlawska}
\bibinfo{author}{\bibfnamefont{J.}~\bibnamefont{Mod{\l}awska}}
  \bibnamefont{and} \bibinfo{author}{\bibfnamefont{A.}~\bibnamefont{Grudka}},
  \bibinfo{journal}{arXiv:0708.0667v1}  (\bibinfo{year}{2007}).

\end{thebibliography}
\end{document}